\documentclass{article}
\usepackage{arxiv}
\usepackage[utf8]{inputenc} 
\usepackage[T1]{fontenc}    
\usepackage{hyperref}
\usepackage{ upgreek }
\usepackage{url}            
\usepackage{booktabs}       
\usepackage{amsfonts}       
\usepackage{nicefrac}       
\usepackage{microtype}      
\usepackage{lipsum}
\usepackage{float}
\usepackage{graphicx}
\usepackage{multirow}
\usepackage{array}
\usepackage{comment}
\usepackage{longtable}
\usepackage{ragged2e}
\usepackage{subfig}
\usepackage{amsmath}
\usepackage{longtable}
\usepackage{ amssymb }
\title{Invariant Scattering Transform for Medical Imaging}

\author{
  Nafisa Labiba Ishrat Huda   \\
  Research and Development Department, Pioneer Alpha,\\
  Dhaka, Bangladesh\\
  \texttt{nafisahuda42@gmail.com} \\
   \And
    Angona Biswas   \\
 Research and Development Department, Pioneer Alpha,\\
  Dhaka, Bangladesh\\
  \texttt{angonabiswas28@gmail.com} \\
   \And
      MD Abdullah Al Nasim \\
   Research and Development Department, Pioneer Alpha,\\
  Dhaka, Bangladesh\\
  \texttt{nasim.abdullah@ieee.org} \\
  \And
    Md. Fahim Rahman \\
   Research and Development Department, Pioneer Alpha,\\
  Dhaka, Bangladesh\\
  \texttt{fahimpranto002gmail.com} \\
  \And
  Shoaib Ahmed \\
   Research and Development Department, Pioneer Alpha,\\
  Dhaka, Bangladesh\\
  \texttt{2010.ahmed.shoaib@gmail.com} \\
}

\begin{document}
\maketitle

\begin{abstract}
Invariant scattering transform introduces new area of research that merges the signal processing with deep learning for computer vision. Nowadays, Deep Learning algorithms are able to solve a variety of problems in medical sector. Medical images are used to detect diseases brain cancer or tumor, Alzheimer’s disease, breast cancer, Parkinson’s disease and many others. During pandemic back in 2020, machine learning and deep learning has played a critical role to detect COVID-19 which included mutation analysis, prediction, diagnosis and decision making. Medical images like X-ray, MRI known as magnetic resonance imaging, CT scans are used for detecting diseases. There is another method in deep learning for medical imaging which is scattering transform. It builds useful signal representation for image classification. It is a wavelet technique; which is impactful for medical image classification problems. This research article discusses scattering transform as the efficient system for medical image analysis where it’s figured by scattering the signal information implemented in a deep convolutional network. A step by step case study is manifested at this research work. 

\end{abstract}

\section{Introduction}

In the field of medical imaging, a professional creates various images for diagnostic purposes. Medical imaging \cite{oreiller2022robust,wang2021fine,guo2022noise,gaudio2023deepfixcx} plays a significant role in modern science. It’s not only includes X-rays \cite{seibert1988x}, MRI \cite{stouffer2022projective}, CT \cite{ohnesorge1999efficient} scans as neuro \cite{kim2015statistical} images but also uses endoscopy, ultrasounds, tactile imaging. Artificial intelligence is enhancing the capability to analyze results in medical imaging technology. Also computer vision is used to diagnose human conditions visually. Manual medical image review procedures have saved countless lives over the years, but that doesn't mean the procedures can't be improved. Machine learning (ML) technology can analyze and learn from vast amounts of data while performing computer vision (CV) analysis on an image to find tiny differences that the human eye cannot detect. Deep Learning (DL) \cite{bharath2017deep} is a form of machine learning (ML) based on artificial neural networks; these technologies have made automated, precise and ultra-fast analysis of medical images a reality. These techniques can achieve high accuracy in analyzing medical images \cite{easley2015image,sahiner2019deep,merone2019computer,al2021operational,kahlessenane2021robust,seibert1988medical} for detection, segmentation, classification and prediction. 

As for image classification, there are two techniques which are supervised and unsupervised. Supervised classification has labeled data images; these are trained data. On the other hand, unsupervised includes unlabeled data that does not have any trained images. It analyzes result by discovering unknown or hidden pattern. To perform an image classification using ML or DL neural networks \cite{haque2023incongruity}, there are some steps to follow. First, image pre-processing is to improve image data so that models can benefit from the data. Image resizing, data augmentation; these are the parts of pre-processing. Then comes feature extraction, it uses for identifying the pattern of the given image. Features can be different to particular classes. This technique helps models to find out the differences between two images. There are also training and testing part where data can be labeled or unlabeled. The last part of this process is classification of the image whether it’s matched with the targeted ones or not. 

On the subject of classification, there are many classifiers such as support vector machine (SVM), K-nearest neighbor (KNN), naïve bayes algorithm, random forest (RB) algorithm. Support vector machine (SVM) demonstrates different classes in a hyper plane in multi-dimensional space. It divides datasets into classes and find out the maximum marginal hyper plane. On the other hand, KNN look at the given data then calculates its distance; find nearest neighbors and count the vote to check the majority of labels.  Random forest (RF) is used for both classification and regression. This algorithm consists of many decision trees. From those trees, it predicts the majority one as a result. 

To use medical images in machine learning and deep learning applications with minimal configuration, a wavelet scattering \cite{raj2021target,shamaei2021wavelet,pham2023classification} enables to obtain low variance features from real time image data. This scattering \cite{bharath2018webrtc} process is acknowledged as scattering transform. Scattering network uses predefined wavelets and scaling filters. It processes data in stages. The outcome of previous stage is the input of next stage. There are three operations in each stage.

\begin{figure}
\centering
\includegraphics[height=2.9cm]{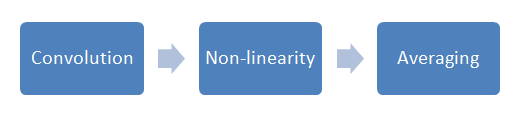}
\caption{Three operations of scattering transform.}
\label{springer fig1.png}
\end{figure}

Fig. \ref{springer fig1.png} is showing the operations of scattering transform network. First, it takes the wavelet transform of the input data with each wavelet filter. Then takes the modulus of the each filtered outputs and average that modulus with the scaling filter. This process is repeated at every node.

\section{Literature Review}
The authors in \cite{minaee2015iris} used scattering transform based features and texture features for iris detection. Not only these two techniques are used; they also applied principal component analysis (PCA) on the extracted features to reduce the dimensionality of the feature. The mean and variance of each scattering images are used as features. Iris database is used by them where it contains 2240 images from 224 different people. An integrated scattering feature for medical image is proposed by the authors of \cite{lan2018integrated}. A corresponding feature demonstration strategy for each integration model is used. Rushi and Yicong applied histogram of compressed scattering coefficients also called HCSCs \cite{lan2016medical}. This proposal took advantages from both compressed scattering coefficients and bag-of-words (BoW) model. For intrapartum fetal heart rate (FHR) variability fractal analysis authors in \cite{chudavcek2013scattering} also used scattering transform. Paper \cite{daliri2012automated} presented automated diagnosis of Alzheimer disease using the scale invariant feature transforms from brain \cite{shah2019brain,al2022brain,hossain2019brain} MR images. Support vector machine (SVM) also applied to the data. The authors of \cite{omer2023lung} applied wavelet scattering transform (WST) \cite{sharma2015left} on CT lung images to detect lung cancer and support vector machine (SVM), kernel nearest neighbor (KNN), wavelet scattering transform, shearlet transform, random forest (RF) as classifiers.

They \cite{gauthier2022parametric} focused on Morlet wavelets and learned the scales, orientations, and aspect ratios of the filters to produce problem-specific parameterizations of scattering transform; also their learned versions of the scattering transform yield significant performance gains in small-sample classification settings over the standard scattering transform. In \cite{bargsten2021attention}, they used scattering transform registration algorithm to align high and low resolution images. A varifold based modeling framework also used by them to compute multiscale data.  They \cite{stouffer2021picoscale} developed a novel network where they used features of a scattering transform for an attention mechanism. An approach \cite{nahak2023fragment} was proposed to classify 17-classes of cardiac arrhythmia using wavelet scattering transform (WST). The paper \cite{ahmad2017mallat} focused on the development of a seizure detection method with the above-stated characteristics where group invariant scattering, novel data representation technique used for feature extraction. They \cite{li2019heart} built a classification scheme that involved a feature extraction and classification stages; they used scattering transform for extracting distinguishes features robust to small deformations in the image. For large dimension of scattering feature matrix, \cite{minaee2017palmprint} used multidimensional scaling (MDS) method to reduce the dimension and the method was compared with the classical dimension reduction method known as principal component analysis (PCA). The paper \cite{szczkesna2022novel} involved the utilization of Support Vector Machine (SVM) classifiers with three non-linear kernels, along with the implementation of k-fold cross-validation to stage detection of Dyslexia in EEG signals using wavelet scattering transform (WST).

In \cite{bikowski2010direct}, they applied a feature learning algorithm known as Wavelet scattering network to retinal fundus images and studied the efficiency of the automatically extracted features therefore for glaucoma detection. \cite{masquelin2021wavelet} projected hybrid framework that fuses the complex shearlet scattering transform (CSST) and a convolutional neural network (CNN) into a single model. A novel motion-compensation method for investigating in vivo myocardium structures in DTI with free-breathing acquisitions was proposed by paper \cite{martinez2022machine}. The paper \cite{nguyen2016integrated} used the wavelet scattering transform (WST)  as a feature extraction technique to obtain features from PPG data and combined it with clinical data for detection of early hypertension stages by applying Early and Late Fusion.  To calculate invariants from the responses to the set of spherical harmonics projected onto 3D kernels in the form of a lightweight Solid Spherical Energy (SSE) CNN (convolutional neural network) \cite{sekhar2021dermoscopic} was applied by them \cite{nguyen2017lbp}. A deep scattering network was used for palm print recognition \cite{rakotomamonjy2014scattering}; scattering network also used in this process to extract features. Fibered confocal fluorescence microscopy (FCFM) imaging technique and scattering features for lung cancer detection \cite {rao2021content}. A hybrid Scattering Coefficients - Bag of Visual Words – Discrete Wavelet Transform (SC-BoVW- DWT) relevance fusion algorithm was proposed for effective content-based medical image retrieval (CBMIR) \cite{kashif2016handcrafted}.

Texture\cite{wang2016texture} descriptor derived from a region covariance matrix of scattering coefficients for detection of glandular structures \cite{raja2020brain} in colon histology images. The procedure \cite{agboola2023wavelet} used features of a wavelet scattering transform to classify signal segments whether it’s chaotic or non-chaotic. The proposed method \cite{berthomier2016venous} of classification determined the level of the brain tumor using the features of the segments generated based on Bayesian fuzzy clustering. They \cite{baharlouei2022detection} proposed a twin branch CNN that learned robust representations by combining deep narrow band features and WST features of the narrow band images to classify vocal cord NBI images into malignant and benign classes. They \cite{ren2021detection} approached a brain tumor classification using a hybrid deep auto encoder with a Bayesian fuzzy clustering-based segmentation. They \cite{andrearczyk2019solid} aimed to analyze the clot texture using the scattering operator which combined wavelet transform convolutions with non-linear modulus and averaging operators. They \cite{raju2018bayesian} applied a novel algorithm which offered immunity to local errors in the underlying deformation field obtained from registration procedures. 

\section{Methodology of study}

\subsection {Wavelet Scattering Transform}

Low variance features from real-time series image data is enabled by wavelet scattering transform (WST) network. WST \cite{rasti2019supervised,pereira2022melanoma} transforms small deformations by separating the variations across different scales. It provides a light representation. There are wavelets, scaling function and input data. There are also a number of users’ particular rotations of the wavelet. Edges from the root to a node are referred as a path. Scattering coefficients are the coefficients convolved with scaling function. These scattering coefficients are the low variance features derived from the data. 


\begin{equation}
    Hx(u) = x  * h(2u)
\end{equation}

\begin{equation}
    Gx(u) = x * g(2u) 
\end{equation}

where h is a low frequency and g is a high frequency.

\subsection {Feature Extraction}

To extract features from data, wavelet Scattering will create and configure the network. Parameters include the size of invariance scale, the number of filter banks and the number of wavelets per octave in each filter bank also set the number of rotation per wavelet. To derive time series and image data features use the object function ScatteringTransform.

\subsection {Invariance Scale}

In the wavelet scattering network, scaling filter plays an important role. This network is invariant to translations up to the invariance scale. Scaling function determines the size of the invariant in time or space.
Time invariance uses for time series data where duration is the invariance scale. The invariance scale does not exceed by the time support of a wavelet. Frequencies are lower than the invariance scale. Image invariance is for image data, the scale specifies in pixels. The scaling function creates a wavelet image scattering network for the image size.

\subsection {Quality Factor}

For the scattering filter banks, quality factors should be set to create a wavelet scattering network. The quality factor for each filter bank is the number of wavelet filter per octave. Large quality factors are not necessary for image data.

\subsection{Case Description with Dataset information}

This section will discuss about the details of different cases. IIT Delhi database \cite{minaee2015iris} contains 2240 iris images captured from 224 different people. The images of 21 people were for validation set; around half of the images were used as training and testing images. In \cite{lan2018integrated}, TCIA-CT database has 675 images of 19 categories. NEMA-CT and EXACT09-CT used for \cite{lan2016medical} where EXACT09 consists of CT scans at the chest, size of 512 x 512. There are total number of 675 CT images in the database. HCSCs methods were applied to both of the databases. The data \cite{daliri2012automated} consists of the scans from 98 normal and 100 scan from Alzheimer disease (AD). The subjects include both men and women. Whole data was for training set and one subject's data was for testing set. In addition, by analyzing the data a method's effectiveness can be increased.

The Bilkent dataset was used by them \cite{raja2020brain} which consists of 72 microscopic images of H\&E stained healthy colon biopsies from 36 patients; where the resolution of images was 480x640 and the dataset was divided into 24 images for training and 48 images for testing. For \cite{sirinukunwattana2015novel}, they used two-dimensional image slices extracted from MRI image stacks. For detecting seizure \cite{ahmad2017mallat}, open access databases featuring multichannel scalp EEG was used; the used dataset is sampled at 256 samples per second and quantized with a 16 bit analog to digital converter where each case contains continuous recordings of the patients and 140 records contained 197 seizures. The testing and training datasets \cite{agboola2023wavelet} were created using 13 dynamical systems of first, second or third order; each system was provided with 1000 created test files, each of which contained 1000 samples. BRATS database for \cite{berthomier2016venous} which had two grades (datasets) of tumor images and every patient image was stored as four different modalities. The dataset \cite{rao2021content} was composed of 103 images taken from 8 healthy volunteers, and 70 images from 7 patients with diagnosed bronchial squamous cell carcinoma. The approached model consisted of two parallel branches where each branch was a ResNet18. The first branch received the NBI images of vocal cord and the second branch received the wavelet scattering transform feature maps of the NBI images of vocal cord. They \cite{lan2018integrated} used data from the National Lung Screening Trial (NLST) in which individuals were screened for lung cancer using low-dose CT (LDCT). Annotated 3700 NLST images contained single solid nodules between 4 and 20 mm in diameter were included in their \cite{deng2019cnn} dataset. A dataset of 15 images of H\&E stained colorectal cancer tissues of 20× optical magnification were used by \cite{bhattacharya2022learning}; where they used a window of size 27 × 27 and stride of 3 to make the patches along spatial dimension (height and width) from the input feature map. In \cite{ren2021detection}, they used the skull removed images (dataset BRATS 2015). \cite{rakotomamonjy2014scattering} used 6000 palm prints sampled from 500 persons. Each palm print was taken under 4 different lights in two different days and preprocessed and had a size of 128x128. Both with synthetic images and real brain image datasets were used in \cite{raju2018bayesian}.

The first dataset contained 1000 synthetic 32×32×32 texture volumes of two classes \cite{nguyen2017lbp}. The second one was a subsample of the American National Lung Screening Trial (NLST) annotated by radiologists which included 485 pulmonary nodules, 244 benign and 241 malignant, from distinct patients in CT. For the work \cite{nguyen2016integrated} was collected at the Guilin People's Hospital in Guilin, China. There were a total of 219 samples from patients admitted to the hospital; the sample had 37\% of subjects diagnosed with NT, 38\% with PHT and 25\% with Stage 1 and Stage 2 hypertension. The approach \cite{masquelin2021wavelet} experimented on two public datasets demonstrate the superiority of our method. The categorized 17 arrhythmia classes were taken from the MIT-BIH arrhythmia database; also they were having 1000 ECG fragments of 45 subjects \cite{nahak2023fragment}.  

\begin{figure}
\centering
\includegraphics[height=5cm]{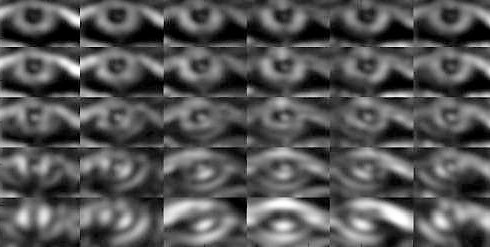}
\caption{Iris images from first layer of scattering transform from \cite{minaee2015iris}.}
\label{iris.jpg}
\end{figure}

\begin{figure}
\centering
\includegraphics[height=4cm]{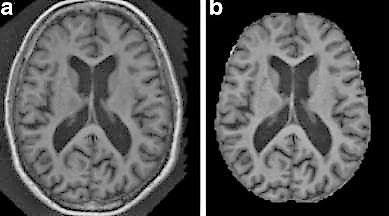}
\caption{MR images extracted by the SIFT method \cite{daliri2012automated}.}
\label{MR.jpg}
\end{figure}

\begin{figure}
\centering
\includegraphics[height=3.0cm]{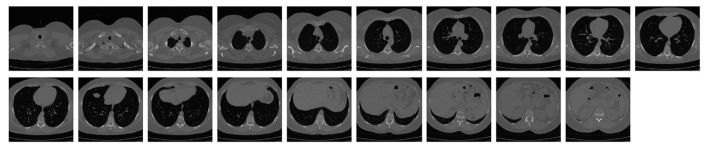}
\caption{CT images from EXACT09-CT database \cite{lan2018integrated}.}
\label{CT.jpg}
\end{figure}

\begin{figure}
\centering
\includegraphics[height=6cm]{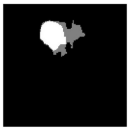}
\caption{tumor images from \cite{berthomier2016venous}.}
\label{brain tumor 38.jpg}
\end{figure}

\begin{figure}
\centering
\includegraphics[height=6cm]{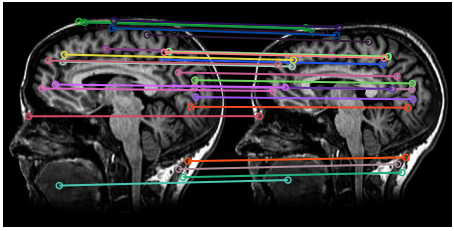}
\caption{medical images from \cite{sirinukunwattana2015novel}.}
\label{medical shearlet 28.jpg}
\end{figure}

\begin{figure}
\centering
\includegraphics[height=6cm]{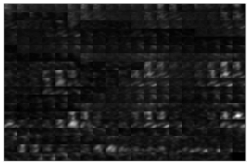}
\caption{palmprint scattering images from \cite{rakotomamonjy2014scattering}.}
\label{palm print 18.jpg}
\end{figure}

\begin{figure}
\centering
\includegraphics[height=6cm]{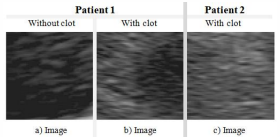}
\caption{clot images from \cite{andrearczyk2019solid}.}
\label{blood clot 39.jpg}
\end{figure}

\section{Result and analysis}

An analysis of some related methods has been presented here.

From Table 1, it is noticed that several studies have been utilized scattering transform method. Others have been conducted different techniques. In paper \cite{minaee2015iris}, they used scattering transform and textural features where they gained best accuracy of 99.2 percent. For each image they \cite{minaee2015iris} applied two levels with a set of filter banks. Minimum distance classifier has been used. There has been two types of integrated scattering data \cite{lan2018integrated} known as data concentration and canonical analysis. To perform scattering transform to a medical image to achieve its translation invariant demonstration, HCSCs method should develop. After that, histogram was derived as feature vector using BoW framework. The proposal \cite{daliri2012automated} used scale-invariant feature transform (SIFT) method for 2-D scan of Magnetic Resonance Imaging (MRI).

From other references, some observations are noted here.
The proposed method \cite{ahmad2017mallat} successfully detected 180 out of 197 seizures (91.4\%) across 24 subjects. This process did not require any pre-processing on the data which was unsupervised and fully automated. CT-images \cite{kashif2016handcrafted} used as a query image and matched with other images in the database. The TCIA-CT database the proposed SC-BoVW-DWT obtained better results than other methods. The proposed method \cite{berthomier2016venous} of tumor classification used the two datasets where the method of tumor segmentation and classification offered better accuracy when compared with the other methods. The PPG signal \cite{agboola2023wavelet} was split into 1000 samples; there were 31.2s segments. The WST features were obtained for each segment. The experiment \cite{raja2020brain} was conducted on 24 Bilkent training images for super pixel features, using 4-fold cross validation; the result showed encouraging segmentation accuracy. They \cite{raju2018bayesian} used three different datasets with different noise levels in the deformation field for registration. 6000 palm prints were sampled from 500 persons from the PolyU palm print database \cite{rakotomamonjy2014scattering}. Each palm print was taken under 4 different lights in two different days; preprocessed and had a size of 128x128, also each image was first divided into blocks of size 32x32. They used half of the images of each person in the dataset for training and the other half for testing and for SVM they were able to get 99.9\% accuracy rate. The recognition accuracy of minimum distance classifier was slightly less than SVM but still high in percentage. The work \cite{bhattacharya2022learning} has showed that detection accuracy of deep convolutional neural networks can be improved by adding handcrafted features to the raw data. The Bayesian fuzzy clustering-based segmentation technique \cite{ren2021detection} for brain tumor classification reached the highest accuracy (98.5 \%) when compared to other existing systems. Their \cite{baharlouei2022detection} results indicated that incorporating wavelet scattering features to improve the generalizing capability of the CNN.

The mentioned SSE-CNN \cite{nguyen2017lbp} largely outperformed the latter > 90\% with N > 2 versus 84.0\% on the synthetic dataset. The analysis \cite{nguyen2016integrated} showed that the wavelet scattering transform (WST) combined with a support vector machine (SVM) can classify norm tension and pre hypertension with an accuracy of 71.42\% and an F1-score of 76\%. They \cite{bikowski2010direct} achieved 98\% on the held-out test set. The WST-based approach \cite{szczkesna2022novel} achieved a high accuracy rate, with an average accuracy of 96.96\% and 97.12\% for dataset 1 and dataset 2 using the RBF kernel. The research \cite{parmar4404532novel} showed the accuracy of 97.4\% and 100\% in diagnosis abnormal retina and DR from normal ones. They also achieved the accuracy of 84.2\% in classifying OCT images to five classes of normal, CSR, MH, AMD and DR which outperformed other state of the art methods with high computational complexity. The classification accuracy of the proposed method \cite{minaee2017palmprint} can reach 98\% or greater than that. The proposed methodology \cite{ahmad2017mallat} successfully detected 180 out of 197 seizures. An overall accuracy of 98.90\% was achieved by them \cite{nahak2023fragment}. This methodology \cite{bargsten2021attention} improved the results of calcium segmentation up to 8.2\% in terms of the Dice coefficient and 24.8\% in terms of the modified Hausdorff distance.

\begin{table}[]
\caption{Performance evaluation of various studies for scattering transform of medical imaging:}
\begin{tabular}{ | p{2cm} | p{3cm} | p{2cm} | p{3cm} | p{1cm} | p{3cm} |}
\hline
Authors                     & Method                                       & Dataset                           & Purpose                                         & Medical image   & Accuracy                                                                                                                  \\ \hline
Shervin et al.\cite{minaee2015iris}       & Scattering transform and textural   features & IIT Delhi iris data               & Iris recognition                                & Iris      & 99.2\%                                                                                                                    \\ \hline
Rushi et al. \cite{lan2018integrated}        & Integrated scattering feature                & TCIA-CT, EXACT09-CT data          & Application to medical retrieval                & CT        &                                                                                                                           \\ \hline
Rushi and Yicong \cite{lan2016medical}    & HCSCs                                        & NEMA, TCIA, EXACT09-CT data & Develop HCSCs for medical image retrieval       & CT        &  NEMA-CT: 3.16\%, 7.37\%,  6.30\% ;\  TCIA-CT: 7.60\%, 10.92\%,  10.53\% \\
\hline
Joakim et al. \cite{chudavcek2013scattering}       & Scattering transform                         & FIGO-TN, FIGO-TP, FIGO-FP         & Fetal heart rate variability fractal   analysis & CTG & FIGO-TN, FIGO-TP (j1=2)                                                                                                   \\ \hline
Mohammad et al. \cite{daliri2012automated}     & Scale invariant feature transform            & OASIS MRI brain data              & Automated diagnosis of Alzheimer disease        & MR        & 86\%                                                                                                                      \\ \hline
Osama and Yoshifumi \cite{omer2023lung} & WST and AI techniques                        & CT scan image data                & Lung cancer detection                           & CT lung  & \begin{tabular}[c]{@{}l@{}}WST+RF: 99.90\%\\    \\ WST+KNN: 95.28\%\\    \\ WST+SVM: 93.24\%\end{tabular}                 \\ \hline
Tameem et al. \cite{adel20173d}       & Scattering transform                         & AGEhIV, OASIS, ADNI data          & Disease classification                          & MRI      & 82.7\%                                                                                                                    \\ \hline
\end{tabular}
\end{table}

\section{Conclusion}

This chapter represents some studies on scattering transform and other methods of medical imaging \cite{suhail2021review,shadekul2023introduction,biswas2023active,biswas2023generative,nasim2021prominence,jahangir2023introduction}.
There are different types of images such as CT scans, MRI \cite{sarker2023case}, CTG, iris and CT lung imaging for diagnosis purposes. SIFT and HOG can be feature descriptors. The minimum classifier is there to match new images with train data. Principal component analysis (PCA) can apply on features to reduce dimensionality. A 3D scattering transform can replace deep learning methods when we won’t have any massive amount of data points. Some methods were used with different classifiers and got best accuracy. A combination of wavelet scattering transform with random forest, k-nearest neighbor and support vector machine bring the best result respectively. But wavelet scattering transform with random forest outperforms all other methodologies. The ability to wavelet scattering transforms (WST) to take image and convert it to signal. It understands the dataset and extract features from it. The results show that this method can perform well for diagnosis, so it can be useful for medical applications.

\bibliographystyle{unsrt}
\bibliography{main}

\begin{thebibliography}{10}

\bibitem{oreiller2022robust}
Valentin Oreiller, Julien Fageot, Vincent Andrearczyk, John~O Prior, and Adrien
  Depeursinge.
\newblock Robust multi-organ nucleus segmentation using a locally rotation
  invariant bispectral u-net.
\newblock In {\em International Conference on Medical Imaging with Deep
  Learning}, pages 929--943. PMLR, 2022.

\bibitem{wang2021fine}
Xiaoqin Wang, Rushi Lan, Huadeng Wang, Zhenbing Liu, and Xiaonan Luo.
\newblock Fine-grained correlation analysis for medical image retrieval.
\newblock {\em Computers \& Electrical Engineering}, 90:106992, 2021.

\bibitem{guo2022noise}
Zhen Guo, Zhiguang Liu, Qihang Zhang, George Barbastathis, and Michael~E
  Glinsky.
\newblock Noise-resilient approach for deep tomographic imaging.
\newblock {\em arXiv preprint arXiv:2211.15456}, 2022.

\bibitem{gaudio2023deepfixcx}
Alex Gaudio, Asim Smailagic, Christos Faloutsos, Shreshta Mohan, Elvin Johnson,
  Yuhao Liu, Pedro Costa, and Aur{\'e}lio Campilho.
\newblock Deepfixcx: Explainable privacy-preserving image compression for
  medical image analysis.
\newblock {\em Wiley Interdisciplinary Reviews: Data Mining and Knowledge
  Discovery}, page e1495, 2023.

\bibitem{seibert1988x}
James~Anthony Seibert and John~M Boone.
\newblock X-ray scatter removal by deconvolution.
\newblock {\em Medical physics}, 15(4):567--575, 1988.

\bibitem{stouffer2022projective}
Kaitlin~M Stouffer, Menno~P Witter, Daniel~J Tward, and Michael~I Miller.
\newblock Projective diffeomorphic mapping of molecular digital pathology with
  tissue mri.
\newblock {\em Communications Engineering}, 1(1):44, 2022.

\bibitem{ohnesorge1999efficient}
Bernd Ohnesorge, Thomas Flohr, and Klaus Klingenbeck-Regn.
\newblock Efficient object scatter correction algorithm for third and fourth
  generation ct scanners.
\newblock {\em European radiology}, 9(3):563--569, 1999.

\bibitem{kim2015statistical}
Won~Hwa Kim, Sathya~N Ravi, Sterling~C Johnson, Ozioma~C Okonkwo, and Vikas
  Singh.
\newblock On statistical analysis of neuroimages with imperfect registration.
\newblock In {\em Proceedings of the IEEE international conference on computer
  vision}, pages 666--674, 2015.

\bibitem{bharath2017deep}
Ramkrishna Bharath and Pachamuthu Rajalakshmi.
\newblock Deep scattering convolution network based features for ultrasonic
  fatty liver tissue characterization.
\newblock In {\em 2017 39th Annual International Conference of the IEEE
  Engineering in Medicine and Biology Society (EMBC)}, pages 1982--1985. IEEE,
  2017.

\bibitem{easley2015image}
Glenn~R Easley, Monica Barbu-McInnis, and Demetrio Labate.
\newblock Image registration using the shearlet transform.
\newblock In {\em Wavelets and sparsity XVI}, volume 9597, pages 125--135.
  SPIE, 2015.

\bibitem{sahiner2019deep}
Berkman Sahiner, Aria Pezeshk, Lubomir~M Hadjiiski, Xiaosong Wang, Karen
  Drukker, Kenny~H Cha, Ronald~M Summers, and Maryellen~L Giger.
\newblock Deep learning in medical imaging and radiation therapy.
\newblock {\em Medical physics}, 46(1):e1--e36, 2019.

\bibitem{merone2019computer}
Mario Merone, Carlo Sansone, and Paolo Soda.
\newblock A computer-aided diagnosis system for hep-2 fluorescence intensity
  classification.
\newblock {\em Artificial intelligence in medicine}, 97:71--78, 2019.

\bibitem{al2021operational}
Ahmed Al-Saffar, A~Zamani, A~Stancombe, and A~Abbosh.
\newblock Operational learning-based boundary estimation in electromagnetic
  medical imaging.
\newblock {\em IEEE Transactions on Antennas and Propagation},
  70(3):2234--2245, 2021.

\bibitem{kahlessenane2021robust}
Fares Kahlessenane, Amine Khaldi, Redouane Kafi, and Salah Euschi.
\newblock A robust blind medical image watermarking approach for telemedicine
  applications.
\newblock {\em Cluster computing}, 24(3):2069--2082, 2021.

\bibitem{seibert1988medical}
JA~Seibert and JM~Boone.
\newblock Medical image scatter suppression by inverse filtering.
\newblock In {\em Medical Imaging II}, volume 914, pages 742--750. SPIE, 1988.

\bibitem{haque2023incongruity}
Md~Aminul Haque~Palash, Akib Khan, Kawsarul Islam, MD~Abdullah Al~Nasim, and
  Ryan~Mohammad Bin~Shahjahan.
\newblock Incongruity detection between bangla news headline and body content
  through graph neural network.
\newblock In {\em The Fourth Industrial Revolution and Beyond: Select
  Proceedings of IC4IR+}, pages 375--387. Springer, 2023.

\bibitem{raj2021target}
Raghu~G Raj, Maxine~R Fox, and Ram~M Narayanan.
\newblock Target classification in synthetic aperture radar images using
  quantized wavelet scattering networks.
\newblock {\em Sensors}, 21(15):4981, 2021.

\bibitem{shamaei2021wavelet}
Amirmohammad Shamaei, Jana Starcukov{\'a}, and Zenon Starcuk~Jr.
\newblock A wavelet scattering convolutional network for magnetic resonance
  spectroscopy signal quantitation.
\newblock In {\em BIOSIGNALS}, pages 268--275, 2021.

\bibitem{pham2023classification}
Tuan~D Pham.
\newblock Classification of motor-imagery tasks using a large eeg dataset by
  fusing classifiers learning on wavelet-scattering features.
\newblock {\em IEEE Transactions on Neural Systems and Rehabilitation
  Engineering}, 31:1097--1107, 2023.

\bibitem{bharath2018webrtc}
R~Bharath and Pachamuthu Rajalakshmi.
\newblock Webrtc based invariant scattering convolution network for automated
  validation of ultrasonic videos for iot enabled tele-sonography.
\newblock In {\em 2018 IEEE 4th World Forum on Internet of Things (WF-IoT)},
  pages 790--795. IEEE, 2018.

\bibitem{minaee2015iris}
Shervin Minaee, AmirAli Abdolrashidi, and Yao Wang.
\newblock Iris recognition using scattering transform and textural features.
\newblock In {\em 2015 IEEE signal processing and signal processing education
  workshop (SP/SPE)}, pages 37--42. IEEE, 2015.

\bibitem{lan2018integrated}
Rushi Lan, Huadeng Wang, Si~Zhong, Zhenbing Liu, and Xiaonan Luo.
\newblock An integrated scattering feature with application to medical image
  retrieval.
\newblock {\em Computers \& Electrical Engineering}, 69:669--675, 2018.

\bibitem{lan2016medical}
Rushi Lan and Yicong Zhou.
\newblock Medical image retrieval via histogram of compressed scattering
  coefficients.
\newblock {\em IEEE journal of biomedical and health informatics},
  21(5):1338--1346, 2016.

\bibitem{chudavcek2013scattering}
V{\'a}clav Chud{\'a}{\v{c}}ek, Joakim And{\'e}n, St{\'e}phane Mallat, Patrice
  Abry, and Muriel Doret.
\newblock Scattering transform for intrapartum fetal heart rate variability
  fractal analysis: a case-control study.
\newblock {\em IEEE Transactions on Biomedical Engineering}, 61(4):1100--1108,
  2013.

\bibitem{daliri2012automated}
Mohammad~Reza Daliri.
\newblock Automated diagnosis of alzheimer disease using the scale-invariant
  feature transforms in magnetic resonance images.
\newblock {\em Journal of medical systems}, 36:995--1000, 2012.

\bibitem{shah2019brain}
Faisal~Muhammad Shah, Tonmoy Hossain, Mohsena Ashraf, Fairuz~Shadmani Shishir,
  MD~Abdullah Al~Nasim, and Md~Hasanul Kabir.
\newblock Brain tumor segmentation techniques on medical images-a review.
\newblock {\em International Journal of Scientific \& Engineering Research},
  10(2):1514--1525, 2019.

\bibitem{al2022brain}
Md~Abdullah Al~Nasim, Abdullah Al~Munem, Maksuda Islam, Md~Aminul~Haque Palash,
  Md~Mahim~Anjum Haque, and Faisal~Muhammad Shah.
\newblock Brain tumor segmentation using enhanced u-net model with empirical
  analysis.
\newblock In {\em 2022 25th International Conference on Computer and
  Information Technology (ICCIT)}, pages 1027--1032. IEEE, 2022.

\bibitem{hossain2019brain}
Tonmoy Hossain, Fairuz~Shadmani Shishir, Mohsena Ashraf, MD~Abdullah Al~Nasim,
  and Faisal~Muhammad Shah.
\newblock Brain tumor detection using convolutional neural network.
\newblock In {\em 2019 1st international conference on advances in science,
  engineering and robotics technology (ICASERT)}, pages 1--6. IEEE, 2019.

\bibitem{omer2023lung}
Osama~A Omer and Yoshifumi Saijo.
\newblock Lung cancer detection using wavelet scattering transform and
  artificial intelligence technique.
\newblock {\em Research Square}, 2023.

\bibitem{sharma2015left}
Upanshu Sharma and Remco Duits.
\newblock Left-invariant evolutions of wavelet transforms on the similitude
  group.
\newblock {\em Applied and Computational Harmonic Analysis}, 39(1):110--137,
  2015.

\bibitem{gauthier2022parametric}
Shanel Gauthier, Benjamin Th{\'e}rien, Laurent Alsene-Racicot, Muawiz
  Chaudhary, Irina Rish, Eugene Belilovsky, Michael Eickenberg, and Guy Wolf.
\newblock Parametric scattering networks.
\newblock In {\em Proceedings of the IEEE/CVF Conference on Computer Vision and
  Pattern Recognition}, pages 5749--5758, 2022.

\bibitem{bargsten2021attention}
Lennart Bargsten, Katharina~A Riedl, Tobias Wissel, Fabian~J Brunner, Klaus
  Schaefers, Michael Grass, Stefan Blankenberg, Moritz Seiffert, and Alexander
  Schlaefer.
\newblock Attention via scattering transforms for segmentation of small
  intravascular ultrasound data sets.
\newblock In {\em Medical Imaging with Deep Learning}, pages 34--47. PMLR,
  2021.

\bibitem{stouffer2021picoscale}
Kaitlin~M Stouffer, Zhenzhen Wang, Eileen Xu, Karl Lee, Paige Lee, Michael~I
  Miller, and Daniel~J Tward.
\newblock From picoscale pathology to decascale disease: Image registration
  with a scattering transform and varifolds for manipulating multiscale data.
\newblock In {\em Multimodal Learning for Clinical Decision Support: 11th
  International Workshop, ML-CDS 2021, Held in Conjunction with MICCAI 2021,
  Strasbourg, France, October 1, 2021, Proceedings 11}, pages 1--11. Springer,
  2021.

\bibitem{nahak2023fragment}
Sudestna Nahak, Akanksha Pathak, and Goutam Saha.
\newblock Fragment-level classification of ecg arrhythmia using wavelet
  scattering transform.
\newblock {\em Expert Systems with Applications}, 224:120019, 2023.

\bibitem{ahmad2017mallat}
Muhammad~Zubair Ahmad, Awais~Mehmood Kamboh, Sajid Saleem, and Amir~Ali Khan.
\newblock Mallat’s scattering transform based anomaly sensing for detection
  of seizures in scalp eeg.
\newblock {\em IEEE Access}, 5:16919--16929, 2017.

\bibitem{li2019heart}
Jinghui Li, Li~Ke, Qiang Du, Xiaodi Ding, Xiangmin Chen, and Danni Wang.
\newblock Heart sound signal classification algorithm: a combination of wavelet
  scattering transform and twin support vector machine.
\newblock {\em IEEE Access}, 7:179339--179348, 2019.

\bibitem{minaee2017palmprint}
Shrevin Minaee and Yao Wang.
\newblock Palmprint recognition using deep scattering network.
\newblock In {\em 2017 IEEE international symposium on circuits and systems
  (ISCAS)}, pages 1--4. IEEE, 2017.

\bibitem{szczkesna2022novel}
Agnieszka Szcz{\k{e}}sna, Dariusz Augustyn, Henryk Josi{\'n}ski, Adam
  {\'S}wito{\'n}ski, Pawe{\l} Kasprowski, and Katarzyna Har{\k{e}}{\.z}lak.
\newblock Novel photoplethysmographic signal analysis via wavelet scattering
  transform.
\newblock In {\em Computational Science--ICCS 2022: 22nd International
  Conference, London, UK, June 21--23, 2022, Proceedings, Part III}, pages
  641--653. Springer, 2022.

\bibitem{bikowski2010direct}
Jutta Bikowski, Kim Knudsen, and Jennifer~L Mueller.
\newblock Direct numerical reconstruction of conductivities in three dimensions
  using scattering transforms.
\newblock {\em Inverse Problems}, 27(1):015002, 2010.

\bibitem{masquelin2021wavelet}
Axel~H Masquelin, Nicholas Cheney, C~Matthew Kinsey, and Jason~HT Bates.
\newblock Wavelet decomposition facilitates training on small datasets for
  medical image classification by deep learning.
\newblock {\em Histochemistry and cell biology}, 155:309--317, 2021.

\bibitem{martinez2022machine}
Erick Martinez-R{\'\i}os, Luis Montesinos, and Mariel Alfaro-Ponce.
\newblock A machine learning approach for hypertension detection based on
  photoplethysmography and clinical data.
\newblock {\em Computers in Biology and Medicine}, 145:105479, 2022.

\bibitem{nguyen2016integrated}
Vu-Lam Nguyen, Ngoc-Son Vu, Hai-Hong Phan, and Philippe-Henri Gosselin.
\newblock An integrated descriptor for texture classification.
\newblock In {\em 2016 23rd International Conference on Pattern Recognition
  (ICPR)}, pages 2006--2011. IEEE, 2016.

\bibitem{sekhar2021dermoscopic}
Kotra Sankar~Raja Sekhar, Tummala~Ranga Babu, Goriparthi Prathibha, Kotra
  Vijay, and Long~Chiau Ming.
\newblock Dermoscopic image classification using cnn with handcrafted features.
\newblock {\em Journal of king Saud University-science}, 33(6):101550, 2021.

\bibitem{nguyen2017lbp}
Vu-Lam Nguyen, Ngoc-Son Vu, Hai-Hong Phan, and Philippe-Henri Gosselin.
\newblock Lbp-and-scatnet-based combined features for efficient texture
  classification.
\newblock {\em Multimedia Tools and Applications}, 76:22425--22444, 2017.

\bibitem{rakotomamonjy2014scattering}
Alain Rakotomamonjy, Caroline Petitjean, Mathieu Sala{\"u}n, and Luc
  Thiberville.
\newblock Scattering features for lung cancer detection in fibered confocal
  fluorescence microscopy images.
\newblock {\em Artificial intelligence in medicine}, 61(2):105--118, 2014.

\bibitem{rao2021content}
R~Varaprasada Rao and T~Jaya~Chandra Prasad.
\newblock Content-based medical image retrieval using a novel hybrid scattering
  coefficients-bag of visual words-dwt relevance fusion.
\newblock {\em Multimedia Tools and Applications}, 80(8):11815--11841, 2021.

\bibitem{kashif2016handcrafted}
Muhammad~Nasim Kashif, Shan E~Ahmed Raza, Korsuk Sirinukunwattana, Muhammmad
  Arif, and Nasir Rajpoot.
\newblock Handcrafted features with convolutional neural networks for detection
  of tumor cells in histology images.
\newblock In {\em 2016 IEEE 13th International Symposium on Biomedical Imaging
  (ISBI)}, pages 1029--1032. IEEE, 2016.

\bibitem{wang2016texture}
Juan Wang, Jiangshe Zhang, and Jie Zhao.
\newblock Texture classification using scattering statistical and cooccurrence
  features.
\newblock {\em Mathematical Problems in Engineering}, 2016, 2016.

\bibitem{raja2020brain}
PM~Siva Raja et~al.
\newblock Brain tumor classification using a hybrid deep autoencoder with
  bayesian fuzzy clustering-based segmentation approach.
\newblock {\em Biocybernetics and Biomedical Engineering}, 40(1):440--453,
  2020.

\bibitem{agboola2023wavelet}
Hafeez~Alani Agboola and Jesuloluwa~Emmanuel Zaccheus.
\newblock Wavelet image scattering based glaucoma detection.
\newblock {\em BMC Biomedical Engineering}, 5(1):1, 2023.

\bibitem{berthomier2016venous}
Thibaud Berthomier, Ali Mansour, Luc Bressollette, Fr{\'e}d{\'e}ric Le~Roy, and
  Dominique Mottier.
\newblock Venous blood clot structure characterization using scattering
  operator.
\newblock In {\em 2016 2nd International Conference on Frontiers of Signal
  Processing (ICFSP)}, pages 73--80. IEEE, 2016.

\bibitem{baharlouei2022detection}
Zahra Baharlouei, Hossein Rabbani, and Gerlind Plonka.
\newblock Detection of retinal abnormalities in oct images using wavelet
  scattering network.
\newblock In {\em 2022 44th Annual International Conference of the IEEE
  Engineering in Medicine \& Biology Society (EMBC)}, pages 3862--3865. IEEE,
  2022.

\bibitem{ren2021detection}
Qingyun Ren, Bingyin Zhou, Liang Tian, and Wei Guo.
\newblock Detection of covid-19 with ct images using hybrid complex shearlet
  scattering networks.
\newblock {\em IEEE Journal of Biomedical and Health Informatics},
  26(1):194--205, 2021.

\bibitem{andrearczyk2019solid}
Vincent Andrearczyk, Valentin Oreiller, Julien Fageot, Xavier Montet, and
  Adrien Depersinge.
\newblock Solid spherical energy (sse) cnns for efficient 3d medical image
  analysis.
\newblock In {\em Proceedings of the 21st Irish Machine Vision and Image
  Processing Conference (IMVIP 2019)}, number CONFERENCE. 28-30 August 2019,
  2019.

\bibitem{raju2018bayesian}
A~Ratna Raju, P~Suresh, and R~Rajeswara Rao.
\newblock Bayesian hcs-based multi-svnn: a classification approach for brain
  tumor segmentation and classification using bayesian fuzzy clustering.
\newblock {\em Biocybernetics and Biomedical Engineering}, 38(3):646--660,
  2018.

\bibitem{rasti2019supervised}
Pejman Rasti, Ali Ahmad, Salma Samiei, Etienne Belin, and David Rousseau.
\newblock Supervised image classification by scattering transform with
  application to weed detection in culture crops of high density.
\newblock {\em Remote Sensing}, 11(3):249, 2019.

\bibitem{pereira2022melanoma}
Pedro~MM Pereira, Lucas~A Thomaz, Luis~MN Tavora, Pedro~AA Assuncao, Rui~M
  Fonseca-Pinto, Rui~Pedro Paiva, and Sergio~MM de~Faria.
\newblock Melanoma classification using light-fields with morlet scattering
  transform and cnn: Surface depth as a valuable tool to increase detection
  rate.
\newblock {\em Medical Image Analysis}, 75:102254, 2022.

\bibitem{sirinukunwattana2015novel}
Korsuk Sirinukunwattana, David~RJ Snead, and Nasir~M Rajpoot.
\newblock A novel texture descriptor for detection of glandular structures in
  colon histology images.
\newblock In {\em Medical Imaging 2015: Digital Pathology}, volume 9420, pages
  186--194. SPIE, 2015.

\bibitem{deng2019cnn}
Zeyu Deng, Lihui Wang, Zixiang Kuai, Qijian Chen, Xinyu Cheng, Feng Yang, Jie
  Yang, and Yuemin Zhu.
\newblock Cnn-based invertible wavelet scattering for the investigation of
  diffusion properties of the in vivo human heart in diffusion tensor imaging.
\newblock {\em arXiv preprint arXiv:1912.07776}, 2019.

\bibitem{bhattacharya2022learning}
Debayan Bhattacharya, Finn Behrendt, Axelle Felicio-Briegel, Veronika Volgger,
  Dennis Eggert, Christian Betz, and Alexander Schlaefer.
\newblock Learning robust representation for laryngeal cancer classification in
  vocal folds from narrow band images.
\newblock In {\em Medical Imaging with Deep Learning}, 2022.

\bibitem{parmar4404532novel}
Shankar Parmar and Chirag Paunwala.
\newblock A novel and efficient primitive stage detection of dyslexia in eeg
  signals using wavelet scattering transform.
\newblock {\em Available at SSRN 4404532}.

\bibitem{adel20173d}
Tameem Adel, Taco Cohen, Matthan Caan, Max Welling, AGEhIV study group,
  Alzheimer's Disease~Neuroimaging Initiative, et~al.
\newblock 3d scattering transforms for disease classification in neuroimaging.
\newblock {\em NeuroImage: Clinical}, 14:506--517, 2017.

\bibitem{suhail2021review}
K~Suhail and D~Brindha.
\newblock A review on various methods for recognition of urine particles using
  digital microscopic images of urine sediments.
\newblock {\em Biomedical Signal Processing and Control}, 68:102806, 2021.

\bibitem{shadekul2023introduction}
SK~Shadekul~Islam, MD~Abdullah Al~Nasim, Ismail Hossain, Md~Azim~Ullah, Kishor
  Datta~Gupta, and Md~Monjur Hossain~Bhuiyan.
\newblock Introduction of medical imaging modalities.
\newblock {\em arXiv e-prints}, pages arXiv--2306, 2023.

\bibitem{biswas2023active}
Angona Biswas, MD~Nasim, Md~Shahin Ali, Ismail Hossain, Dr~Md~Azim Ullah, and
  Sajedul Talukder.
\newblock Active learning on medical image.
\newblock {\em arXiv preprint arXiv:2306.01827}, 2023.

\bibitem{biswas2023generative}
Angona Biswas, MD~Nasim, Al~Imran, Anika~Tabassum Sejuty, Fabliha Fairooz, Sai
  Puppala, and Sajedul Talukder.
\newblock Generative adversarial networks for data augmentation.
\newblock {\em arXiv preprint arXiv:2306.02019}, 2023.

\bibitem{nasim2021prominence}
MD~Nasim, Aditi Dhali, Faria Afrin, Noshin~Tasnim Zaman, and Nazmul Karim.
\newblock The prominence of artificial intelligence in covid-19.
\newblock {\em arXiv preprint arXiv:2111.09537}, 2021.

\bibitem{jahangir2023introduction}
Md~Zihad~Bin Jahangir, Ruksat Hossain, Riadul Islam, MD~Nasim, Md~Mahim~Anjum
  Haque, Md~Jahangir Alam, and Sajedul Talukder.
\newblock Introduction to medical imaging informatics.
\newblock {\em arXiv preprint arXiv:2306.00421}, 2023.

\bibitem{sarker2023case}
Shuvra Sarker, Angona Biswas, MD~Nasim, Md~Shahin Ali, Sai Puppala, and Sajedul
  Talukder.
\newblock Case studies on x-ray imaging, mri and nuclear imaging.
\newblock {\em arXiv preprint arXiv:2306.02055}, 2023.

\end{thebibliography}
\end{document}